\documentstyle[12pt]{article}
\begin{document}
\hfill{NCKU-HEP-98-06}\par
\vskip 0.5cm
\begin{center}
{\large {\bf Small-$x$ evolution with $Q$ dependence and unitarity}}
\vskip 1.0cm
Hsiang-nan Li
\vskip 0.5cm
Department of Physics, National Cheng-Kung University, \par
Tainan, Taiwan, Republic of China
\end{center}
\vskip 1.0cm

\vskip 1.0cm

\centerline{\bf Abstract}
\vskip 0.3cm

We propose a modified Balitsky-Fadin-Kuraev-Lipatov equation from the
viewpoint of the resummation technique, which contains an intrinsic
dependence on momentum transfer $Q$, and satisfies the unitarity
bound. The idea is to relax the strong rapidity ordering and to restrict
phase space for real gluon emissions in the evaluation of the BFKL kernel.
It is shown that the power-law rise of the gluon distribution function
with the small Bjorken variable $x$ turns into a logarithmic rise, and
that the predictions for the proton structure function $F_2(x,Q^2)$ are
consistent with the HERA data.

\newpage

\centerline{\large \bf 1. Introduction}
\vskip 0.5cm

It is known that the Balitsky-Fadin-Kuraev-Lipatov (BFKL) equation
\cite{BFKL} sums leading logarithms $\ln (1/x)$, $x$ being the Bjorken
variable, which are produced from the reggeon ladder diagrams with the
rung gluons obeying the strong rapidity ordering. This equation predicts a
rise of the gluon distribution function and thus a rise the proton structure
function $F_2$ involved in deep inelastic scattering (DIS) at small $x$,
which have been confirmed by the recent HERA data \cite{H1}. However,
some controversies remain unsolved.

Since the BFKL equation is independent of momentum transfer $Q$, its
predictions are insensitive to the variation of $Q$. On the contrary, the
experimental data for $F_2(x,Q^2)$ exhibit a stronger $Q$ dependence
\cite{H1}: The rise at low $x$ is slower for smaller $Q$, corresponding to
so-called soft pomeron exchanges. The rise obtained from the BFKL equation
is always larger, which corresponds to hard pomeron exchanges. To explain
the data, the Dokshitzer-Gribov-Lipatov-Altarelli-Parisi (DGLAP) equation
\cite{AP}, which sums single logarithms $\ln Q$ produced from strong
transverse momentum ordering, is then employed. The relevant splitting
function $P_{gg}$ contains a term $1/x$, which also gives a rise at
small $x$. An alternative solution \cite{KMS} is the
Ciafaloni-Catani-Fiorani-Marchesini (CCFM) equation, which embodies the
DGLAP and BFKL equations \cite{CCFM}. For this unified evolution equation,
the $Q$ dependence comes from the summation of the ladder diagrams with
strong angular ordering.

On the other hand, the increase of the gluon distribution function at 
$x\to 0$ predicted by the BFKL equation is power-like, such that $F_2$ and
the DIS cross section $\sigma_{\rm tot}$ rise as a power of $x$, {\it i.e.},
as a power of $s$ for $x=Q^2/s$, $s$ being the total energy. This behavior
does not satisfy the Froissart bound $\sigma_{\rm tot} \le {\rm const.}
\times\ln^2 s$, and violates unitarity. Hence, the BFKL equation can not be
the final theory for small $x$ physics. Though it has been expected that the
inclusions of next-to-leading $\ln(1/x)$ \cite{FL} and of higher-twist
effects from the exchange of multiple pomerons \cite{MP} may soften the
BFKL rise, the attempts have not yet led to a concrete conclusion.

Recently, we have proposed a modified BFKL equation from the
viewpoint of the resummation technique \cite{CS}, which involves an
intrinsic $Q$ dependence \cite{L4}. This $Q$ dependence was introduced
by cutting off the longitudinal component of the real gluon momentum at a
scale of order $Q$, when evaluating the BFKL kernel. The motivation is
that it is unlikely for a real gluon to carry an infinite momentum. The
gluon distribution function derived from this $Q$-dependent evolution
equation gives predictions of $F_2$, which are in good agreement with the
data. Unfortunately, the predictions still rise as a power of $x$. In
\cite{L5} we observed that relaxing the strong rapidity ordering for real
gluon emissions results in a destructive correction, and the power-law rise
of the gluon distribution function turns into a logarithmic rise. The
unitarity bound is thus satisfied.

In this letter we shall combine the above two modifications to derive
a new small-$x$ evolution equation, which depends on $Q$ and satisfies the
unitarity bound at the same time. We demonstrate that the predictions for
$F_2$ are consistent with the current data for $x > 10^{-4}$, and that the
logarithmic rise of $F_2$ turns on as $x < 10^{-4}$.
\vskip 1.0cm

\centerline{\large \bf 2. The formalism}
\vskip 0.5cm

We refer the detailed derivation of the the BFKL equation using the
resummation technique to \cite{L3}. Here we summarize only the basic
idea. The unintegrated gluon distribution function $F(x,k_T)$, describing
the probability of a parton carrying a longitudinal momentum fraction $x$
and transverse momenta ${\bf k}_T$, is defined by
\begin{eqnarray}
F(x,k_T)&=&\frac{1}{p^+}\int\frac{dy^-}{2\pi}\int\frac{d^2y_T}{4\pi}
e^{-i(xp^+y^--{\bf k}_T\cdot {\bf y}_T)}
\nonumber \\
& &\times\frac{1}{2}\sum_\sigma\langle p,\sigma|
F^+_\mu(y^-,y_T)F^{\mu+}(0)|p,\sigma\rangle\;,
\label{deg}
\end{eqnarray}
in the axial gauge $n\cdot A=0$, $n$ being a gauge vector with $n^2\not=0$.
The ket $|p,\sigma\rangle$ denotes the incoming proton with the light-like
momentum $p^\mu=p^+\delta^{\mu +}$ and spin $\sigma$. An average over color
is understood. $F^+_\mu$ is the field tensor.

For a fixed parton momentum $k^+$, $F$ varies with $p^+$
implicitly through $x=k^+/p^+$, and we have $-xdF/dx=p^+dF/dp^+$.
That is, the differentiation of $F$ with respect to $x$ is related to the
differentiation with respect to $p^+$. Because of the scale invariance in
$n$ of the gluon propagator $-iN^{\mu\nu}(l)/l^2$ with
\begin{equation}
N^{\mu\nu}=g^{\mu\nu}-\frac{n^\mu l^\nu+n^\nu l^\mu}
{n\cdot l}+n^2\frac{l^\mu l^\nu}{(n\cdot l)^2}\;,
\label{gp}
\end{equation}
$F$ must depend on $p$ and $n$ via the ratio $(p\cdot n)^2/n^2$.
Hence, there exists a chain rule relating $p^+d/dp^+$ to $d/dn$,
\begin{eqnarray}
p^+\frac{d}{dp^+}F=-\frac{n^2}{v\cdot n}v_\beta\frac{d}{dn_\beta}F\;,
\label{cph}
\end{eqnarray}
$v_\beta=\delta_{\beta +}$ being a vector along $p$. The operator
$d/dn_\beta$ applies to a gluon propagator, giving
\begin{equation}
\frac{d}{dn_\beta}N^{\nu\nu'}=
-\frac{1}{n\cdot l}(l^\nu N^{\beta\nu'}+l^{\nu'} N^{\nu\beta})\;.
\label{dgp}
\end{equation}
The loop momentum $l^\nu$ ($l^{\nu'}$) contracts with the vertex the
differentiated gluon attaches, which is then replaced by a special vertex
${\hat v}_\beta=n^2v_\beta/(v\cdot nn\cdot l)$. This special vertex can be
read off from the combination of Eqs.~(\ref{cph}) and (\ref{dgp}).

Summing the diagrams with different differentiated gluons, and employing
the Ward identity for the contraction of $l^\nu$ ($l^{\nu'}$), those
diagrams embedding the special vertices cancel by pairs, leaving the one in
which the special vertex moves to the outer end of the parton line \cite{CS}.
We then obtain the derivative,
\begin{equation}
-x\frac{d}{dx}F(x,k_T)=2{\bar F}(x,k_T)\;,
\label{df}
\end{equation}
where the new function $\bar F$ contains one special vertex \cite{L3}. The
coefficient 2 comes from the equality of the new functions with the special
vertex on either of the two parton lines.

Next we factorize the new function into the convolution of the subdiagram
involving the special vertex with the original gluon distribution function.
The resultant factorization formula will lead to the small-$x$ evolution
equation. The kinematic approximation, {\it i.e.}, the rapidity ordering of
radiative gluons, is specified only at the stage of computing the
subdiagram, which is then identified as the evolution kernel. Therefore,
in our approach it is easier to handle the approximation applied to the
evolution kernel, and to investigate its effect on the behavior of the
gluon distribution function.

The only important region of the loop momentum $l$ flowing through the
special vertex is soft, in which the subdiagram containing the special
vertex can be factorized. Including the color factor, the factorization
formula is written as \cite{L4,L5}
\begin{eqnarray}
{\bar F}(x,k_T)&=&iN_cg^2\int\frac{d^{4}l}{(2\pi)^4}N^{\nu\beta}(l)
\frac{{\hat v}_\beta v_\nu}{v\cdot l}
\left[2\pi i\delta(l^2)F(x+l^+/p^+,|{\bf k}_T+{\bf l}_T|)\right.
\nonumber \\
& &\left.+\frac{\theta(k_T^2-l_T^2)}{l^2}F(x,k_T)\right]\;.
\label{kf1}
\end{eqnarray}
The first term in the brackets corresponds to the real gluon emission,
where $F(x+l^+/p^+,|{\bf k}_T+{\bf l}_T|)$ implies that the parton coming
out of the proton carries the momentum components $xp^++l^+$ and
${\bf k}_T+{\bf l}_T$ in order to radiate a real gluon of momentum $l$. The
second term corresponds to the virtual gluon emission, where the $\theta$
function sets the upper bound of $l_T$ to $k_T$ to ensure a soft momentum
flow.

The integration over $l^-$ gives
\begin{eqnarray}
{\bar F}(x,k_T)&=&\frac{{\bar\alpha}_s}{2}\int\frac{d^{2}l_T}{\pi}
\left[\int_0^{\infty} dl^+\frac{2l^+ n^2}{(2n^-l^{+2}+n^+l_T^2)^2}
\right.
\nonumber \\
& &\left.\times F(x+l^+/p^+,|{\bf k}_T+{\bf l}_T|)
-\frac{\theta(k_T^2-l_T^2)}{l_T^2}F(x,k_T)\right]\;,
\label{kf2}
\end{eqnarray}
with ${\bar \alpha}_s=N_c\alpha_s/\pi$, $N_c=3$ being the number of colors.
We shall show that Eq.~(\ref{kf2}) reduces to the conventional BFKL
equation, the modified BFKL equation with the $Q$ dependence, the modified
BFKL equation with unitarity, and the new evolution equation with both the
modifications by removing the applied approximations gradually.

To derive the conventional BFKL equation, we simply assume the strong
rapidity ordering, namely, approximate $F(x+l^+/p^+)$ by its dominant value
$F(x)$. Performing the integration over $l^+$ to infinity, and substituting
${\bar F}$ into Eq.~(\ref{df}), we arrive at
\begin{eqnarray}
\frac{dF(x,k_T)}{d\ln(1/x)}=
{\bar \alpha}_s\int\frac{d^{2}l_T}{\pi l_T^2}
\left[F(x,|{\bf k}_T+{\bf l}_T|)-\theta(k_T^2-l_T^2)F(x,k_T)\right]\;,
\label{bfkl}
\end{eqnarray}
which is the BFKL equation adopted in \cite{KMS}.

However, the vanishing of $F(x+l^+/p^+)$ at large momentum fraction
constrains $l^+$ to go to infinity. To render the approximation
$F(x+l^+/p^+)\approx F(x)$ more reasonable, we truncate $l^+$ at the scale
$Q/\sqrt{2}$ for the real gluon emission in Eq.~(\ref{kf2}), and obtain
the $Q$-dependent BFKL equation,
\begin{eqnarray}
\frac{dF(x,k_T)}{d\ln(1/x)}&=&
{\bar \alpha}_s\int\frac{d^{2}l_T}{\pi l_T^2}
\left[F(x,|{\bf k}_T+{\bf l}_T|)-\theta(k_T^2-l_T^2)F(x,k_T)\right]
\nonumber \\
& &-{\bar \alpha}_s\int\frac{d^{2}l_T}{\pi}
\frac{F(x,|{\bf k}_T+{\bf l}_T|)}{l_T^2+Q^2}\;,
\label{bfklq}
\end{eqnarray}
for the choice $n=(1,1,{\bf 0})$. The extra term compared to the
conventional BFKL equation comes from the upper bound of $l^+$. It is
trivial to find that Eq.(\ref{bfklq}) approaches Eq.~(\ref{bfkl}) as
$Q\to \infty$. This correction, being negative, moderates the BFKL rise at
low $Q$. Equation (\ref{bfklq}) has been solved in \cite{L4}, and its
predictions of the structure function $F_2$ match the data
for various $x$ and $Q^2$.

As stated in the Introduction, the conventional BFKL equation gives a
power-law rise for the gluon distribution function at $x\to 0$, which
violates unitarity. It has been pointed out that the assumption of the
strong rapidity ordering is the cause for the unitarity violation \cite{L5}.
For most values of $l^+$, $F(x+l^+/p^+)$ is much smaller than $F(x)$. Hence,
replacing the former by the latter in the whole integration range of $l^+$
overestimates the real gluon contribution, which is responsible for the
rise. To derive an evolution equation with unitarity, we employ
Eq.~(\ref{kf2}) directly without applying the approximation.
Reexpress $F(x+l^+/p^+,|{\bf k}_T+{\bf l}_T|)$ as
\begin{eqnarray}
F(x,|{\bf k}_T+{\bf l}_T|)
+\left[F(x+l^+/p^+,|{\bf k}_T+{\bf l}_T|)
-F(x,|{\bf k}_T+{\bf l}_T|)\right]\;,
\label{corr}
\end{eqnarray}
in Eq.~(\ref{kf2}), where the first term, combined with the virtual gluon
contribution, leads to the conventional BFKL equation, and the terms in the
brackets are the correction from relaxing the strong rapidity ordering.
We then derive 
\begin{eqnarray}
\frac{dF(x,k_T)}{d\ln(1/x)}&=&
{\bar\alpha}_s\int\frac{d^{2}l_T}{\pi l_T^2}
\left[F(x,|{\bf k}_T+{\bf l}_T|)
-\theta(k_T^2-l_T^2)F(x,k_T)\right]
\nonumber \\
& &+{\bar\alpha}_s\int\frac{d^{2}l_T}{\pi}
\int_0^{\infty} dl^+\frac{4l^+}{(2l^{+2}+l_T^2)^2}
\nonumber \\
& &\times\left[F(x+l^+/p^+,|{\bf k}_T+{\bf l}_T|)
-F(x,|{\bf k}_T+{\bf l}_T|)\right]\;.
\label{mkf1}
\end{eqnarray}
The above equation has been studied in \cite{L5}, and the BFKL power-law
rise was found to be moderated into a logarithmic rise at small $x$.

\vskip 1.0cm
\centerline{\large \bf 3. The new equation}

\vskip 0.5cm

The $Q$-dependent BFKL equation (\ref{bfklq}), though phenomenologically
successful, still gives a power-law rise of the gluon distribution function.
Hinted by Eq.~(\ref{mkf1}), we attempt to combine the above two
modifications by switching the upper bound of $l^+$ from $\infty$ to
$Q/\sqrt{2}$ in Eq.~(\ref{kf2}), and obtain
\begin{eqnarray}
\frac{dF(x,k_T)}{d\ln(1/x)}&=&{\bar \alpha}_s\int\frac{d^{2}l_T}{\pi}
\left[\int_0^1 dy\frac{2y Q^2}{(y^2Q^2+l_T^2)^2}
F(x+y\sqrt{x},|{\bf k}_T+{\bf l}_T|)\right.
\nonumber \\
& &\left.-\frac{\theta(k_T^2-l_T^2)}{l_T^2}F(x,k_T)\right]\;,
\label{bfklu}
\end{eqnarray}
where the variable change $l^+=y\sqrt{x}p^+$ has been adopted, and
the upper bound of $y$ is determined by the kinematic relation
$Q\approx \sqrt{2x}p^+$. Equation (\ref{bfklu}) is the new small-$x$
evolution equation we shall investigate in more details below.

In order to simplify the analysis, the $\theta$ function for the virtual
gluon emission is replaced by $\theta(Q_0^2-l_T^2)$ \cite{L4}, where the
parameter $Q_0$ can be determined from data fitting. This
simplification is reasonable, because the virtual gluon contribution only
plays the role of a soft regulator for the real gluon emission, and setting
the cutoff of $l_T$ to $Q_0$ serves the same purpose. We then Fourier
transform Eq.~(\ref{bfklu}) into the $b$ space conjugate to $k_T$ with
Eq.~(\ref{corr}) inserted, deriving
\begin{eqnarray}
\frac{d{\tilde F}(x,b)}{d\ln(1/x)}&=&S(b,Q){\tilde F}(x,b)
\nonumber \\
& &+2{\bar \alpha}_s(1/b)Qb\int_0^1 dy K_1(yQb)
[{\tilde F}(x+y\sqrt{x},b)-{\tilde F}(x,b)]\;,
\nonumber \\
& &
\label{bfb}
\end{eqnarray}
where 
\begin{equation}
S(b,Q)=-2{\bar\alpha}_s(1/b)\left[\ln(Q_0 b)+\gamma-\ln 2+K_0(Qb)\right]
\label{es}
\end{equation}
comes from the combination of the first term in Eq.~(\ref{corr}) and the
virtual gluon emission term. Note that $S$ is the evolution kernel for
Eq.~(\ref{bfklq}) in $b$ space. $K_0$ and $K_1$ are the Bessel functions,
and $\gamma$ the Euler constant. The argument of $\alpha_s$ has been set to
the natural scale $1/b$.

An initial condition ${\tilde F}(x_0,b)={\tilde F}^{(0)}(x_0,b)$ must be
assumed when solving Eq.~(\ref{bfb}), $x_0$ being the initial momentum
fraction. For instance, a ``flat" gluon distribution function \cite{KMS}
\begin{equation}
{\tilde F}^{(0)}(x,b)=3N_g(1-x)^5\exp(-Q_0^2b^2/4)\;,
\end{equation}
for $x \ge x_0$, $N_g$ being a normalization constant,
has been proposed. Therefore, the initial
function ${\tilde F}^{(0)}(x+y\sqrt{x},b)$ should be substituted for
${\tilde F}(x+y\sqrt{x},b)$ in Eq.~(\ref{bfb}) as $x+y\sqrt{x}> x_0$.

Before solving Eq.~(\ref{bfb}), we extract the behavior
of ${\tilde F}$ analytically.
Inserting a guess ${\tilde F}\propto x^{-\lambda}$ into Eq.~(\ref{bfb}),
$\lambda$ being a parameter, we obtain
\begin{equation}
\lambda=S+2{\bar \alpha}_sQb\int_0^1 dy K_1(yQb)
\left[\left(\frac{x+y\sqrt{x}}{x}\right)^{-\lambda}-1\right]\;.
\label{tb}
\end{equation}
It can be numerically verified that a solution of $\lambda$, 
$0< \lambda < S$, exists for $x<x_0$. That is, ${\tilde F}$ increases as 
a power of $x$, consistent with the results from the conventional and 
$Q$-dependent BFKL equations. While the correction term, {\it i.e.}, the
second term on the right-hand side of Eq.~(\ref{tb}), diverges as $x\to 0$,
and no solution of $\lambda$ is allowed, implying that ${\tilde F}$ can not 
maintain the power-law rise at extremely small $x$. We then substitute
another guess ${\tilde F}|_{x\to 0}\propto \ln(1/x)$ with a milder 
rise into Eq.~(\ref{bfb}). In this case the correction term, increasing as
$\ln^{1.2}(1/x)$, only slightly dominates over the first term
$S{\tilde F}\propto \ln(1/x)$ (the derivative term
$-xd{\tilde F}/dx\propto 1$ is negligible), and Eq.~(\ref{bfb}) holds
approximately. At last, we assume ${\tilde F}|_{x\to 0}\propto$ const. as
a test. It is easy to find that the first term
becomes dominant, and the correction term and the derivative term vanish,
{\it i.e.}, no const.$\not=0$ exists. These observations indicate that
${\tilde F}$ should increase as $\ln(1/x)$ at most when $x$ approaches zero. 
In conclusion, the new evolution equation predicts a rapid power-like
rise of ${\tilde F}$ for $x< x_0$ and a milder logarithmic rise at $x\to 0$.

The structure function $F_2$ is written, in terms of the
$k_T$-factorization theorem \cite{J}, as
\begin{equation}
F_2(x,Q^2)=\int_x^1 \frac{d\xi}{\xi}\int_0^{p_c} \frac{d^2k_T}{\pi}
H(x/\xi,k_T,Q)F(\xi,k_T)\;,
\label{f2}
\end{equation}
$p_c$ being the upper bound of $k_T$ which will be specified later.
The hard scattering subamplitude $H$ denotes the contribution from the quark
box diagrams, where both the incoming photon and gluon are off shell by
$-Q^2$ and $-k_T^2$, respectively. $H$ has been computed in \cite{L4},
and its expression is given by
\begin{eqnarray}
H(z,k_T,Q)&=&\sum_q e_q^2\frac{\alpha_s}{2\pi}z
\Biggl\{\left[z^2+(1-z)^2-2z(1-2z)\frac{k_T^2}{Q^2}
+2z^2\frac{k_T^4}{Q^4}\right]
\nonumber \\
& &\times
\frac{1}{\sqrt{1-4z^2k_T^2/Q^2}}
\ln\frac{1+\sqrt{1-4z^2k_T^2/Q^2}}{1-\sqrt{1-4z^2k_T^2/Q^2}}-2\Biggr\}\;,
\end{eqnarray}
with $e_q$ the electric charge of the quark $q$.
To require a meaningful $H$, the upper bound of
$k_T$ in Eq.~(\ref{f2}) is set to
\begin{equation}
p_c=\min\left(Q,\frac{\xi}{2x}Q\right)\;.
\end{equation}

We choose $x_0=0.1$ and $Q_0=0.4$ GeV, and evaluate Eq.~(\ref{f2}) for
$Q^2=15$ GeV$^2$. Fitting the results to the corresponding data \cite{H1},
the normalization constant is determined to be $N_g=1.207$. $N_g$ for
other $Q^2$ are then extracted by requiring that the gluon density,
defined by
\begin{equation}
xg(x,Q^2)=\int_0^Q\frac{d^2 k_T}{\pi}F(x,k_T)\;,
\end{equation}
has a fixed normalization $\int_0^1 xg dx$. $F_2$ for
$Q^2=8.5$, 12, and 20 GeV$^2$ are computed, and the results
along with the data \cite{H1} are displayed in Fig.~1.
For comparision, we also present the results from the conventional and
$Q$-dependent BFKL equations \cite{L4}. It is found that the
shape of the curves from the conventional BFKL equation is almost
independent of $Q$, and thus the match with the data is not very
satisfactory. The curves from the $Q$-dependent BFKL equation, exhibiting
smaller slopes for lower $Q$, match
the data. However, they increase rapidly at $x\to 0$, and violate the
unitarity bound. The predictions from the new evolution equation also agree
with the data well. The curves have a steeper rise at
a larger $Q$, which is the consequence of the cutoff at the scale of order
$Q$. Because of $F(\xi,k_T)\le \ln(1/\xi)$ at $\xi \to 0$, $F_2$ rises as
$\ln^2(1/x)$ at most as indicated by the $\xi$ integration in
Eq.~(\ref{f2}), and thus satisfies the unitarity bound. It is observed
that the ascent of the curves does not speed up as obtained from the
conventional and $Q$-dependent BFKL equations at $x< 10^{-4}$. This is the
consequence of relaxing the strong rapidity ordering.

\vskip 1.0cm

\centerline{\large \bf 4. Conclusion}
\vskip 0.5cm

In this letter we have combined the modifications to the conventional
BFKL equation obtained in our previous studies, which are based on
the resummation formalism. The resultant small-$x$ evolution equation 
possesses the intrinsic $Q$ dependence and satisfies the unitarity
requirement. The predictions of the structure
function $F_2$ are in agreement with the HERA data for various $x$ and
$Q^2$. We have also observed that the power-law rise turns into a
logarithmic rise as $x< 10^{-4}$.

We emphasize that the cutoff of order $Q$ for the longitudinal momentum in
the real gluon emission is phenomenologically motivated in this work.
In the derivation of a new unified evolution equation for both
intermediate and small $x$, we shall demonstrate that the scale
$Q$ can be introduced in a rigorous way. This subject will be published
elsewhere \cite{LL2}.

This work was supported by the National Science Council of R.O.C. under the
Grant No. NSC-87-2112-M-006-018.

\newpage
\centerline{\large \bf Figure Caption}
\vskip 0.5cm

\noindent
{\bf FIG. 1.} The dependence of $F_2$ on $x$ for
$Q^2=8.5$, 12, 15, and 20 GeV$^2$ from the conventional BFKL equation
(dotted lines), from the $Q$-dependent BFKL equation, and from the
new evolution equation (solid lines).


\newpage
\begin{thebibliography}{99}
\bibitem{BFKL} E.A. Kuraev, L.N. Lipatov and V.S. Fadin, Sov. Phys. JETP 
{\bf 45} (1977) 199; Ya.Ya. Balitsky and L.N. Lipatov, Sov. J. Nucl. Phys. 
{\bf 28} (1978) 822; L.N. Lipatov, Sov. Phys. JETP {\bf 63} (1986) 904.
\bibitem{H1} ZEUS Collaboration, M. Derrick {\it et al.}, Z. Phys. C 
{\bf 65} (1995) 379; H1 Collaboration, T. Ahmed {\it et al.}, Nucl. Phys. 
{\bf B439} (1995) 471.
\bibitem{AP} V.N. Gribov and L.N. Lipatov, Sov. J. Nucl. Phys. {\bf 15}
(1972) 428; G. Altarelli and G. Parisi, Nucl. Phys. {\bf B126} (1977) 298;
Yu.L. Dokshitzer, Sov. Phys. JETP {\bf 46} (1977) 641.
\bibitem{KMS} J. Kwieci\'nski, A.D. Martin, and P.J. Sutton, Phys. Rev.
D {\bf 53} (1996) 6094.
\bibitem{CCFM} M. Ciafaloni, Nucl. Phys. {\bf B296} (1988) 49;
S. Catani, F. Fiorani, and G. Marchesini, Phys. Lett. B {\bf 234}
(1990) 339; Nucl. Phys. {\bf B336} (1990) 18; G. Marchesini, Nucl. Phys. 
{\bf B445} (1995) 49.
\bibitem{FL} V.S. Fadin and L.N. Lipatov, Nucl. Phys. {\bf B406}
(1993) 259; A.R. White, Phys. Lett. B {\bf 334} (1994) 87; C. Corian\'o
and A.R. White, Nucl. Phys. {\bf B451} (1995) 231.
\bibitem{MP} A.H. Mueller and B. Patel, Nucl. Phys. {\bf B425} (1994) 471.
\bibitem{CS} J.C. Collins and D.E. Soper, Nucl. Phys. {\bf B193}
(1981) 381; H-n. Li, Phys. Rev. D {\bf 55} (1997) 105.
\bibitem{L4} H-n. Li, Phys. Lett. B {\bf 416} (1998) 98; J.L. Lim and
H-n. Li, Report No. hep-ph/9712408.
\bibitem{L5} H-n. Li, Report No. hep-ph/9709236, to appear in Phys. Lett. B.
\bibitem{L3} H-n. Li, Phys. Lett. B {\bf 405} (1997) 347; Report No.
hep-ph/9703328.
\bibitem{J} T. Jaroszewicz, Acta. Phys. Pol. B {\bf 11} (1980) 965;
S. Catani, M. Ciafaloni, and F. Hautmann, Phys. Lett. B {\bf 242}
(1990) 97; Nucl. Phys. {\bf B366} (1991) 657;
S. Catani and F. Hautmann, Nucl. Phys. {\bf B427} (1994) 475.
\bibitem{LL2} J.L. Lim and H-n. Li, in preparation.
\end{thebibliography}
\end{document}